# Analytical form of current-voltage characteristic of parallel-plane, cylindrical and spherical ionization chambers with homogeneous ionization

**Dimitar G STOYANOV**

Faculty of Engineering and Pedagogy in Sliven, Technical University of Sofia
59, Bourgasko Shaussee Blvd, 8800 Sliven, BULGARIA

E-mail: dgstoyanov@abv.bg

**Abstract.** The elementary processes taking place in the formation of charged particles and their flow in parallel-plane, cylindrical and spherical geometry cases of ionization chamber are considered. On the basic of particles and charges balance a differential equation describing the distribution of current densities in the ionization chamber volume is obtained. As a result of the differential equation solution an analytical form of the current-voltage characteristic of a ionization chamber with homogeneous ionization is obtained. For the parallel-plane case the comparison with experimental data is performed.

## 1. Introduction

The ionization chamber is a simple and reliable device for measuring the characteristics of beams of particles and radiations. This refers to either beams of radioactive emission or X-rays, synchrotron radiation and beams of electrons and protons in different accelerators. The ionization chamber has been turned into an irreplaceable tool in such investigations because of which it is necessary the properties and the parameters of the ionization chamber to be got to know in different working regimes.

The ionization chamber represents a volume filled with gas with two metallic electrodes available in it. The radioactive emission passing through the volume causes a partial ionization of the gas molecules and generates charged particles in it. The presence of a potential difference between the metallic electrodes causes the current running through the gas volume whose current carriers are the ionization products.

The objective of this article is the obtaining of current-voltage characteristic of parallel-plane, cylindrical and spherical ionization chambers with homogeneous ionization.

This problem is not a new one. The ionization chamber is the device by means of which the radioactivity is investigated [1] up to the discovery of the Geiger-Müller counter. Attempts for the analytical describing of current-voltage characteristic have been done in the right direction [1, 2], but a solution is not found up to now [3]. Besides, the achievements in the experiment and the theory of the physicochemical processes in plasma during last decades [4] provide a good basis for re-examining the solutions of some old problems from the point of view of the modern concepts [5].

## 2. Elementary processes and balances of particles

During the currency of radioactive emission **R** through the gas volume an ionization process of the gas molecules $A_2$ takes place according the reaction [1, 2, 4]:

$$A_2 + R \xrightarrow{\nu_i} A_2^+ + e + R \tag{1}$$

In the process of ionization positive gas ions and electrons are uniformly generated in the volume possessing the following rate

$$\left(\frac{\delta n_+}{\delta t}\right)_{ion} = \left(\frac{\delta n_e}{\delta t}\right)_{ion} = +\nu_i \cdot N = I_i \tag{2}$$

where $n_+$ is the concentration of the positive gas ions;

$n_e$ is the concentration of electrons in the gas volume;

$\nu_i$ is the frequency of ionization according to reaction (1);

N is the concentration of the neutral gas molecules;

$I_i$ is the ionization rate per unit of volume.

The process (1) is the mechanism through which charged particles are generated in the volume. Thus these charged particles could be lost mainly by two mechanisms:

- Neutralization of the metallic electrodes, when the charged particles drop onto their surface. At low voltages between the metallic electrodes the charged particles clashes with the metallic electrodes do not cause an emission of new electrons from them. The neutralization of the charged particles upon the metallic electrodes is the mechanism for the current closure of the external circuit through the gas volume.
- Volume recombination. During the interaction between a positive ion and an electron in the gas volume they can be neutralized according the reaction [1, 2, 4]:

$$A_2^+ + e \xrightarrow{\beta} A_2 \tag{3}$$

At volume recombination the charged particles are neutralized with the rate

$$\left(\frac{\delta n_+}{\delta t}\right)_{rec} = \left(\frac{\delta n_e}{\delta t}\right)_{rec} = -\beta.n_+.n_e \tag{4}$$

where $\beta$ is the coefficient of two-particle recombination.

When a potential difference is applied between the metallic electrodes an electric field appears inside the volume which causes the appearance of directed flows of charged particles, i.e. current with densities as following:

$$\vec{j}_e = -e.n_e.\vec{V}_{de} = -e.n_e.\mu_e.\vec{E}, \tag{5}$$

$$\vec{j}_+ = e.n_+.\vec{V}_{d+} = e.n_+.\mu_+.\vec{E}, \tag{6}$$

where $\vec{V}_{de}$ is the drift velocity of electrons in the gas;

$\vec{V}_{d+}$ is the drift velocity of positive ions in the gas;

$\mu_e$ is the mobility of electrons in the gas;

$\mu_+$ is the mobility of positive ions in the gas.

Here we assume that there are small gradients of electrons and positive ions concentrations in the volume, wherefore the diffusion flows are insignificant [4].

Then for the balance of charged particles concentrations (a law for conservation the particles number) we write [4]:

$$\frac{\partial n_e}{\partial t} + \nabla.\left(\frac{\vec{j}_e}{-e}\right) = I_i - \beta.n_e.n_+, \tag{7}$$

$$\frac{\partial n_+}{\partial t} + \nabla.\left(\frac{\vec{j}_+}{e}\right) = I_i - \beta.n_e.n_+. \tag{8}$$

In respect to the density of the electric charge **k** and its current $\vec{j}$ the following relations are valid:

$$k = -e.n_e + e.n_+, \tag{9}$$

$$\vec{j} = \vec{j}_e + \vec{j}_+, \tag{10}$$

$$\frac{\partial k}{\partial t} + \nabla.\vec{j} = 0. \tag{11}$$

The last equation is the law for conservation of electric charge.

In the stationary case the partial derivatives of time in the balances are zeroes.

These are the basic elementary processes with the participation of the charged particles and the concentrations balances of the particles and the electric charge. It is necessary Gauss' law of electric-field strength $\vec{E}$ to be added to this system of equations:

$$\nabla \cdot \vec{E} = \frac{k}{\varepsilon_0} = \frac{-e.n_e + e.n_+}{\varepsilon_0} = 0. \tag{12}$$

When considering the ionization chamber irradiated with relatively weak radioactive emission the frequency of ionization is low and the concentrations of the charged particles are small, and they will not change the electric field created by both metallic electrodes, which means that the right side of equation (12) will be nullified [4].

**3. A plane case**

**3.1.** *Geometry and equations*

We will consider farther on a plane case of ionization chamber. Thus both metallic electrodes account for parallel metallic planes situated by distance **d** apart.

The plates have area **S** ($d^2 \ll S$). We assume that the electric field is concentrated merely in the volume between the plates and that it is homogeneous.

For describing the spatial coordinates we choose a coordinate system with **OX**-axis perpendicular to the plates and **OY**- and **OZ**-axes forming a plane parallel to the plates.

Let the cathode have coordinate $x_k=0$ and the anode have coordinate $x_a=d$. At such configuration of the electrodes the vectors of the electric-field strength and the motion rates of charged particles will be directed parallel to **OX**-axis. The vector of electric-field strength is pointed from the anode to the cathode. The electrons move in the direction from the cathode to the anode, and they are absorbed by the anode. The positive ions move in the direction from the anode to the cathode, and they are neutralized by the cathode.

If we express from equations (5) and (6) the concentrations of the charged particles by their currents and mobilities, from the balance of charged particles concentration we obtain [4]:

$$\frac{dj_e}{dx} = e.I_i - e.\beta \cdot \frac{j_e}{e.\mu_e.E} \cdot \frac{j_+}{e.\mu_+.E}, \tag{13}$$

$$\frac{dj_+}{dx} = -e.I_i + e.\beta \cdot \frac{j_e}{e.\mu_e.E} \cdot \frac{j_+}{e.\mu_+.E}, \tag{14}$$

$$\frac{dj}{dx} = \frac{d(j_e + j_+)}{dx} = 0. \tag{15}$$

From equation (15) follows that in the plane case **j** is constant in the gas volume. Whereas the densities of current of electrons (5) and ions (6) could be changed in the gas volume, their sum (10) is compulsory constant.

As a first step it is necessary the solutions of equations (13), (14) and (15) considered as a system to be found. We are looking for a solution satisfying the following boundary conditions over the electrodes [4]:

- cathode does not emit electrons:

$$\mathbf{j_e(x = 0) = 0} \tag{16}$$

- cathode neutralizes the positive ions falling upon it:

$$\mathbf{j_+(x = 0) = j} \tag{17}$$

- anode absorbs the electrons falling upon it:

$$\mathbf{j_e(x = d) = j} \tag{18}$$

- anode does not emit positive ions:

$$\mathbf{j_+(x = d) = 0} \tag{19}$$

We have chosen to find firstly the function $\mathbf{j_e(x)}$. Expressing the current density of the positive ions through the current density of the electrons we obtain:

$$\frac{dj_e}{dx} = e.I_i - \frac{\beta}{e.\mu_e.\mu_+.E^2} \cdot j_e.(j - j_e) \tag{20}$$

Thus we have got an equation that will be mainly studied farther on.

## 3.2. *Regime: No recombination*

We consider this case because it will allow us to study the extreme case when either there is no recombination or the recombination is slightly small, e.g. at very great electric field.

Actually, in this case in the right side of equation (20) remains only the first term which does not depend on **x**.

The solution is simple:

$$j_e(x) = e.I_i.x = e.I_i.d.\frac{x}{d} = j_s.\frac{x}{d} \tag{21}$$

Whence using equation (18) we will obtain

$$j = e.I_i.d = j_s. \tag{22}$$

The function $j_e(x)$ (21) is a solution of equation (20) such that the right side of equation (20) has maximum magnitude (if a recombination is available the right side has a smaller value than that).

Translated into physics language this means that all generated charges in the chamber volume are directed to the electrodes and they reach them. That is the reason the current that is available at this regime to be maximum, and we will call it a ***current of saturation*** $j_s$.

### 3.3 *Regime: With recombination*

When solving the equation (20) in the common case in order to simplify the notations we make it dimensionless to $j_e$ and **x**.

$$z = \frac{x}{d} \in [0,1], \tag{23}$$

$$f(z) = \frac{j_e}{j} \in [0,1]. \tag{24}$$

The function **f(z)** shows how the electron component of current in the chamber volume changes during certain regime of working. Besides, **1-f(z)** will give us an information how the ionic component of current in the volume will change.

Using (23) and (24) equation (20) could be transformed in the following

$$\frac{df}{dz} = \frac{e.I_i.d}{j} - \frac{\beta.d.j}{e.\mu_e.\mu_+.E^2}.f.(1-f) \tag{25}$$

Besides, introducing the substitution

$$4.E_1^2 = \frac{\beta.d.j_s}{e.\mu_e.\mu_+} \tag{26}$$

and using (22) we can carry out (25) in the form

$$\frac{df}{dz} = \frac{j_s}{j} - \frac{4.E_1^2}{E^2}.\frac{j}{j_s}.f.(1-f). \tag{27}$$

This equation (27) should be set in a standard form. For the purpose we get

$$\frac{df}{dz} = \frac{j_s}{j} - \frac{E_1^2}{E^2}\frac{j}{j_s} + \frac{4.E_1^2}{E^2}\frac{j}{j_s}\left(f - \frac{1}{2}\right)^2 = a_0 + a_2.\left(f - \frac{1}{2}\right)^2 \qquad (28)$$

where $\quad a_0 = \dfrac{j_s}{j} - \dfrac{E_1^2}{E^2}\dfrac{j}{j_s},$ \hfill (29)

and

$$a_2 = \frac{4.E_1^2}{E^2}.\frac{j}{j_s}. \qquad (30)$$

According [6] the solution of (28) gives the following

$$\frac{1}{\sqrt{a_0.a_2}}.\arctan\left[\sqrt{\frac{a_2}{a_0}}.\left(f - \frac{1}{2}\right)\right] = z - C. \qquad (31)$$

After taking into account the boundary conditions over cathode and anode, and after certain transformations, is obtained

$$f(z) = \frac{1}{2} + \sqrt{\frac{a_0}{a_2}}.\tan\left[\sqrt{a_0.a_2}.\left(z - \frac{1}{2}\right)\right] \qquad (32)$$

In order to satisfy the boundary conditions is necessary and sufficient

$$\sqrt{\frac{a_0}{a_2}}.\tan\left[\sqrt{a_0.a_2}.\left(\frac{1}{2}\right)\right] = \frac{1}{2} \qquad (33)$$

This equation represents one transcendent equation for the relation between $a_0$ and $a_2$, but in this case it plays the role of an analytical form of the ionization chamber current-voltage characteristic.

### 3.4. *Analysis of the solution*

Using solution (32) we present the right side of equation (27) graphically in figure 1. As it is evident (see figure 1.) a great dynamic change of the electron current particularly close to the electrodes is available. Really, close to the electrodes one of the current components is nullified and because of this the lost of charged particles through recombination is small there. In figure 1. is also seen that at weak electric fields the recombination consumes very strongly the generated charged particles inside the volume. This indicates that the differential equation (20) is written in a correct form.

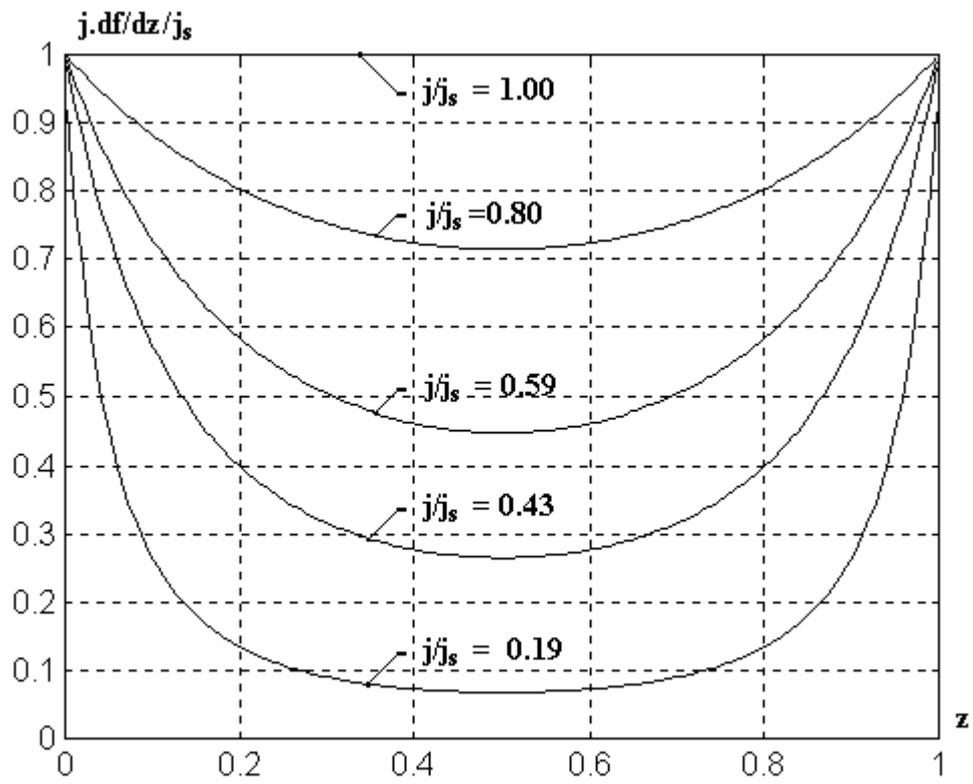

**Figure.1. The right side of equation (27) in relative units.**

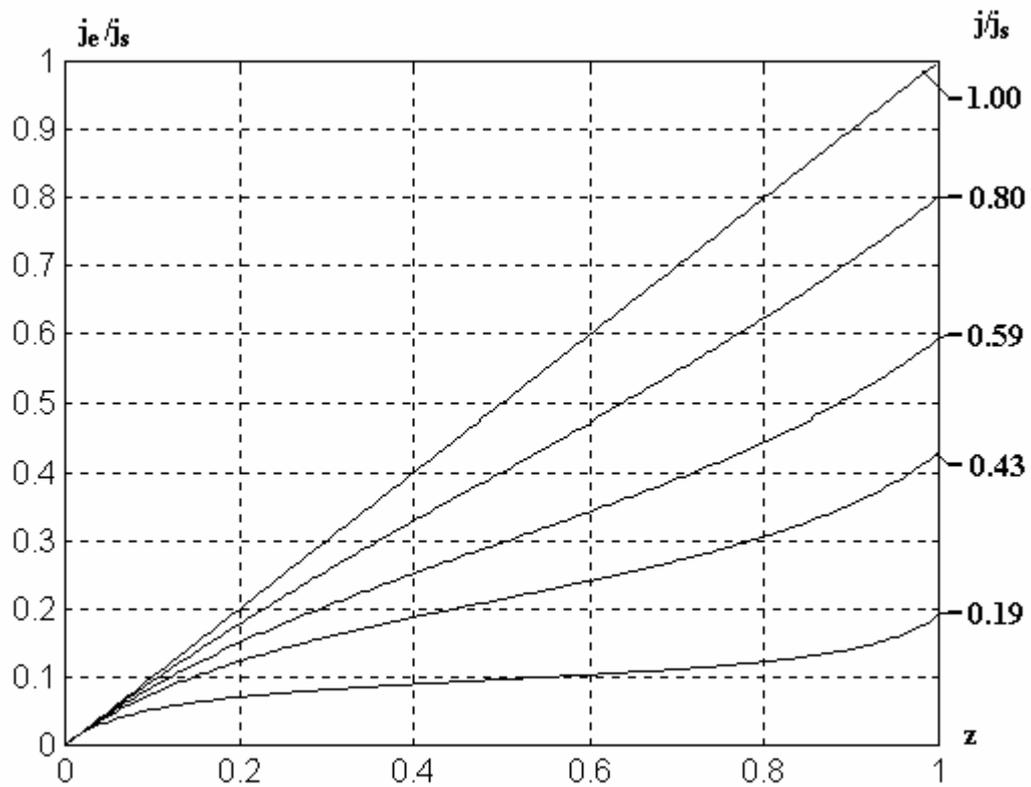

**Figure 2. A curve of set of solutions of (32) at different rate of recombination.**

As a consequence of it the obtained solution (32) has the following graphical form (figure 2).

Finally equation (33) contains the current-voltage characteristic of the parallel-plane ionization chamber. Replacing the constants by their equals from (29) and (30), and making some transformations, we obtain

$$\sqrt{\dfrac{\dfrac{E^2}{E_1^2}-\dfrac{j^2}{j_s^2}}{\dfrac{j^2}{j_s^2}}} \cdot \tan\left[\dfrac{\sqrt{\dfrac{E^2}{E_1^2}-\dfrac{j^2}{j_s^2}}}{\dfrac{E^2}{E_1^2}}\right] = 1 \qquad (34)$$

Equation (34) is transcendent and could be solved numerically. The graphical dependence between the current density and the electric-field strength (the potential difference between both electrodes, respectively) is represented in figure 3.

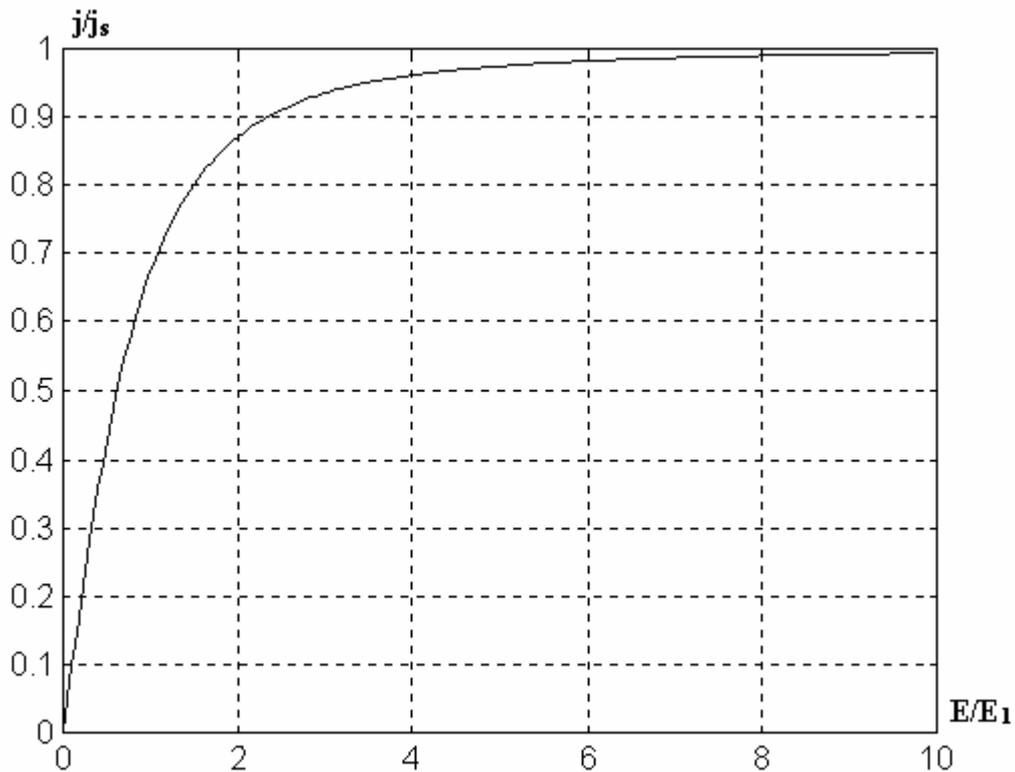

**Figure 3. A current-voltage characteristic of plane ionization chamber**

In table 1 some values satisfying (34) could be read.

As it is evident (figure 3.) with the increase of the electric-field strength the current through the ionization chamber increases as it trends towards the current of saturation. At strong electric fields the current-voltage characteristic could be approximately defined from (34) as:

**Table 1. A current-voltage characteristic of ionization chamber.**

| $\dfrac{E}{E_1}$ | $\dfrac{j}{j_s}$ |
|:---:|:---:|
| 0.00 | 0.0000 |
| 0.10 | 0.0990 |
| 0.20 | 0.1931 |
| 0.30 | 0.2798 |
| 0.40 | 0.3584 |
| 0.50 | 0.4286 |
| 0.60 | 0.4909 |
| 0.80 | 0.5941 |
| 1.00 | 0.6763 |
| 1.20 | 0.7347 |
| 1.40 | 0.7818 |
| 1.60 | 0.8185 |
| 1.80 | 0.8474 |
| 2.00 | 0.8703 |
| 2.50 | 0.9100 |
| 3.00 | 0.9344 |
| 4.00 | 0.9612 |
| 6.00 | 0.9821 |
| 8.00 | 0.9898 |
| 10.00 | 0.9934 |

$$\frac{j}{j_s} \cong 1 - \frac{2}{3}\cdot\left(\frac{E_1}{E}\right)^2 + \frac{4}{5}\cdot\left(\frac{E_1}{E}\right)^4 \tag{35}$$

When the electric fields are weak the recombination consumes strongly the charged particles generated in the volume, and that is the reason the current to be small. In this case the current-voltage characteristic could be approximately defined from (34) as:

$$\frac{j}{j_s} \cong \frac{E}{E_1}\cdot\left[1 - \frac{\pi^2}{8}\cdot\left(\frac{E}{E_1}\right)^2\right] \tag{36}$$

**3.5. *Comparison with the experiment***

In order to be made a comparison with the experiment is necessary the published yet articles containing data for a large range of currents (from 0 to 100 % of the current of saturation) through a parallel-plane ionization chamber and applied voltages to be chosen. These data should be with a great precision in the whole range of conditions.

Therefore the reference [7] was noticed in which the creation of a parallel-plane ionization chamber (with $d = 0.3\,cm$), allocated for medical irradiations, is reported. In this work the results from the measurement of the current $i$ through the ionization chamber are given for series of four air pressures in the chamber volume and for four intensities of ions beam $C^{6+}\,(290\,MeV/u)$, as for each series the measurement are done for at least 15 magnitudes of voltages applied to the chamber. Especially useful for the measurements precision in [7] is turned out the application of the current $i_{cont}$ of another ionization chamber for the control of beam intensity and for the correction of the obtained results according this current. In this case the control ionization chamber is working in a regime of saturation and neither the gas pressure nor the applied voltage to it is changing.

Here we will use only the results from the series with a change in the working pressure of the gas in the chamber at one and the same intensity of ions beam. For these series the dependence between the relative effectiveness $f = i/i_{cont}$ of the investigated ionization chamber in comparison with the control one and the applied voltage $U$ is represented graphically in [7].

We suppose that at a constant intensity of the ionizing irradiation

$$\frac{j}{j_s} = \frac{f}{f_{sat}}, \qquad (37)$$

and

$$\frac{E}{E_1} = \frac{U}{U_1}. \qquad (38)$$

The new magnitudes $f_{sat}$ and $U_1$ are constants within the framework of one series (for one and the same gas pressure) but their values are various for the different series. In this work they are obtained as a result of a selection and a fitting.

Table 2. Values of $f_{sat}$ and $U_1$ at different pressures of the gas in the chamber volume

| P, Torr | $f_{sat}$ | $U_1$, V |
|---|---|---|
| 759.6 | 0.780 | 151.0 |
| 202.2 | 0.213 | 23.3 |
| 111.2 | 0.109 | 12.1 |
| 50.2 | 0.049 | 5.3 |

Using (37), (38) and the data represented in table 2 the data from [7] are shown in figure 4.

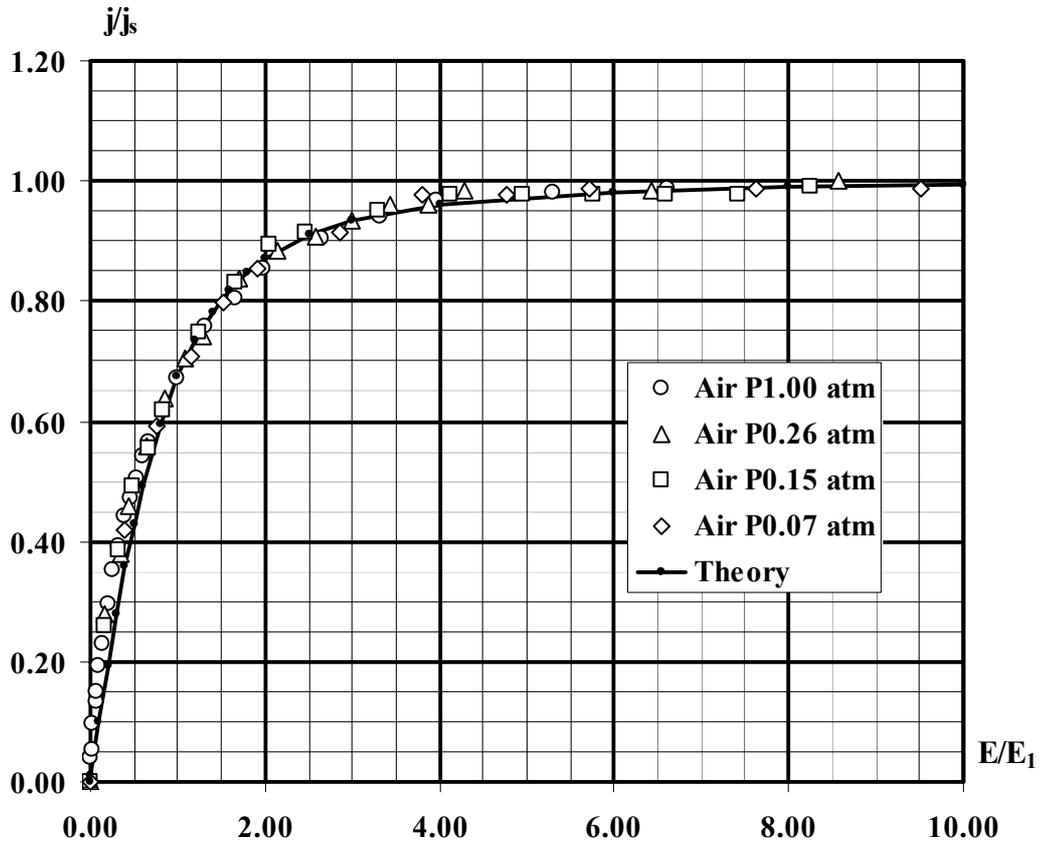

Figure 4. A comparison between the theoretical curve and the experimental points from [7].

As it is evident from figure 4, in the investigated range of values of the electric field strength, the theoretical curve moves either close or between the experimental results. Therefore we may say that as a dynamics of the dependence course and as an agreement between the values there is a good accordance between the theoretical curve and the experimental results.

## 4. A cylindrical and spherical case

### 4.1. *Geometry and equations*

Cylindrical and spherical cases of an ionization chamber working in a stationary regime will be considered further on. In stationary working regime the partial derivatives per time in the balances are zeros.

This means that the equipotential surfaces of the electric field and the form of the electrodes, which are also equipotential surfaces, possess cylindrical and spherical symmetry.

If the electrodes are coaxial cylinders with radii of the cathode and the anode $r_k$ and $r_a$ ($r_k < r_a$), respectively, and a length $L$ we will have a cylindrical symmetry.

If the electrodes are concentric spheres with radii of the cathode and the anode $r_k$ and $r_a$ ($r_k < r_a$), respectively, we will have a spherical symmetry.

We suppose that the electric field is concentrated merely in the volume between the electrodes and submits to (12).

For the description of the spatial coordinates we choose a coordinate system with the corresponding type of symmetry as the radial unit vector is pointed perpendicular to the equipotential surfaces of the electric field and the metallic electrodes in direction from the cathode to the anode (inside – outside).

At so chosen coordinate system the vectors of electric field strength and the velocities of movements of the charged particles will have directions parallel to the radial unit vector. The vector of the electric field strength is directed from the anode to the cathode. The electrons will move in direction from the cathode to the anode, and will be absorbed from the anode. The positive ions move in direction from the anode to the cathode and are neutralized by the cathode.

Therefore, in the operator of divergence present in (7), (8), (11) and (12) only radial partial derivative will remain.

If we express the concentrations of the charged particles by their currents and mobilities from (5) and (6) from the balance of the charged particles concentrations (7), (8) and (11) we could get

$$\frac{1}{r^n} \cdot \frac{\partial}{\partial r}\left(r^n \cdot j_e\right) = q.I_i - \beta \cdot \frac{j_e}{\mu_e.E} \cdot \frac{j_+}{q.\mu_+.E}, \quad (39)$$

$$\frac{1}{r^n} \cdot \frac{\partial}{\partial r}\left(r^n \cdot j_+\right) = -q.I_i + \beta \cdot \frac{j_e}{\mu_e.E} \cdot \frac{j_+}{q.\mu_+.E}, \quad (40)$$

$$\frac{1}{r^n} \cdot \frac{\partial}{\partial r}\left(r^n \cdot j\right) = \frac{1}{r^n} \cdot \frac{\partial}{\partial r}\left[r^n \cdot (j_e + j_+)\right] = 0. \quad (41)$$

Here at **n = 1** we have a cylindrical symmetry, while at **n = 2** the symmetry is spherical.

As a next step it is necessary the solutions of equations (39), (40) and (41) considered as a system to be found. For the solving the following boundary conditions on the electrodes are taken into account:

- the cathode does not emit electrons

$$j_e(r_k) = 0, \tag{42}$$

- the cathode neutralizes the positive ions falling on it

$$j_+(r_k) = j, \tag{43}$$

- the anode absorbs the electrons falling on it

$$j_e(r_a) = j, \tag{44}$$

- the anode does not emit positive ions

$$j_+(r_a) = 0. \tag{45}$$

We choose to find the equation and the solution of $j_e(r)$. Expressing the current density of the positive ions by the current density of the electrons we get:

$$\frac{1}{r^n} \cdot \frac{\partial}{\partial r}(r^n \cdot j_e) = q \cdot I_i - \frac{\beta}{q \cdot \mu_e \cdot \mu_+ \cdot E^2} \cdot j_e \cdot (j - j_e). \tag{46}$$

Thus we got the equation that will be examined farther on.

The solving of the equation will be done as firstly we will make changes of the variables of the field and current densities.

From (12) follows that

$$e = r^n \cdot E = \text{const}_1. \tag{47}$$

From (11) follows that

$$i = r^n . j = \text{const}_2 . \tag{48}$$

The magnitudes defined in (47) and (48) do not depend on the spatial coordinate **r** and therefore they are pointed as constants.

Besides let

$$i_e = r^n . j_e . \tag{49}$$

For the work further it is necessary in (46) the variables connected with the spatial derivatives to be changed.

Let us have a sufficiently small "volume"

$$dV = r^n . dr . \tag{50}$$

Hence for the "volume" between the cathode and the present equipotential surface we will get

$$V = \int_{r_k}^{r} r^n . dr . \tag{51}$$

And for the "volume" between the cathode and the anode we will get

$$V_1 = \int_{r_k}^{r_a} r^n . dr . \tag{52}$$

Therefore, we can introduce and the dimensionless "volume" η

$$\eta = \frac{V}{V_1} \in [0,1]. \tag{53}$$

As for the sufficiently small "volume" $d\eta$ we will get

$$d\eta = \frac{dV}{V_1}$$

Thus after making the spatial coordinates dimensionless and applying the substitution (49) the equation (46) gets the following form

$$\frac{d}{d\eta}(i_e) = q.I_i.V_1 - \frac{\beta.V_1}{q.\mu_e.\mu_+} \cdot \frac{i_e.(i-i_e)}{e^2}. \tag{54}$$

### 4.2. *Regime: No Recombination*

Firstly we will consider this case when either there is no recombination or it is slightingly small, e.g. at very strong electric field.

At such a regime in the right side of equation (54) only the first term remains, not depending on $\eta$.

The solution is

$$i_e(\eta) = q.I_i.V_1.\eta = i_s.\eta \tag{55}$$

Using (44) we will get for the current through the anode

$$i = q.I_i.V_1 = i_s. \tag{56}$$

At this regime all electric charges generated in the chamber volume are directed towards the electrodes and they reach them. Therefore the current running at such a regime will be maximum, and we will call it a *current of saturation* $i_s$.

### 4.3 *Regime: With recombination*

During the solving of (54) in order to simplify the record we will make it dimensionless toward $i_e$.

$$f(\eta) = \frac{i_e}{i} \in [0,1]. \tag{57}$$

The function $f(\eta)$ shows how the current electron component changes in the vessel volume at certain working regime. Besides the function $1-f(\eta)$ gives us information, how the current ionic component changes through the volume.

Using (56) and (57) equation (54) could be transformed in the form

$$\frac{df}{d\eta} = \frac{i_s}{i} - \frac{\beta \cdot V_1 \cdot i}{q \cdot \mu_e \cdot \mu_+ \cdot e^2} \cdot f \cdot (1-f). \tag{58}$$

Putting the substitution

$$4 \cdot e_1^2 = \frac{\beta \cdot V_1 \cdot i_s}{q \cdot \mu_e \cdot \mu_+}, \tag{59}$$

we can get from (58)

$$\frac{df}{d\eta} = \frac{i_s}{i} - \frac{4 \cdot e_1^2}{e^2} \cdot \frac{i}{i_s} \cdot f \cdot (1-f). \tag{60}$$

The equation (60) should be set in a standard form. For this purpose we get

$$\frac{df}{d\eta} = \frac{i_s}{i} - \frac{e_1^2}{e^2} \frac{i}{i_s} + \frac{4 \cdot e_1^2}{e^2} \frac{i}{i_s} \left(f - \frac{1}{2}\right)^2 = a_0 + a_2 \cdot \left(f - \frac{1}{2}\right)^2, \tag{61}$$

where:   $-a_0 = \dfrac{i_s}{i} - \dfrac{e_1^2}{e^2} \dfrac{i}{i_s},$ (62)

and

$-a_2 = \dfrac{4 \cdot e_1^2}{e^2} \cdot \dfrac{i}{i_s}.$ (63)

The solving of (61) in accordance with [6] gives

$$\frac{1}{\sqrt{a_0 \cdot a_2}} \cdot \arctan\left[\sqrt{\frac{a_2}{a_0}} \cdot \left(f - \frac{1}{2}\right)\right] = \eta - C. \tag{64}$$

After taking into account the boundary conditions between the cathode and the anode and after certain transformations we get

$$f(\eta) = \frac{1}{2} + \sqrt{\frac{a_0}{a_2}} \cdot \tan\left[\sqrt{a_0 \cdot a_2} \cdot \left(\eta - \frac{1}{2}\right)\right]. \tag{65}$$

In order to be satisfied the boundary conditions it is necessary and sufficient

$$\sqrt{\frac{a_0}{a_2}} \cdot \tan\left[\sqrt{a_0 \cdot a_2} \cdot \left(\frac{1}{2}\right)\right] = \frac{1}{2}. \tag{66}$$

This equation is a transcendent one for the relation between $a_0$ and $a_2$, but in this case it plays the role of an analytical form of the current-voltage characteristic of the ionization chamber.

### 4.4. *Analysis of the solution*

The current-voltage characteristic of the ionization chamber is hidden in (66). If we substitute the constants with their equals from (62) and (63), and after certain transformations we get

$$\sqrt{\frac{\frac{e^2}{e_1^2} - \frac{i^2}{i_s^2}}{\frac{i^2}{i_s^2}}} \cdot \tan\left[\sqrt{\frac{\frac{e^2}{e_1^2} - \frac{i^2}{i_s^2}}{\frac{e^2}{e_1^2}}}\right] = 1. \tag{67}$$

Using equation (67) we represent graphically the solution of (65) (Figure 5).

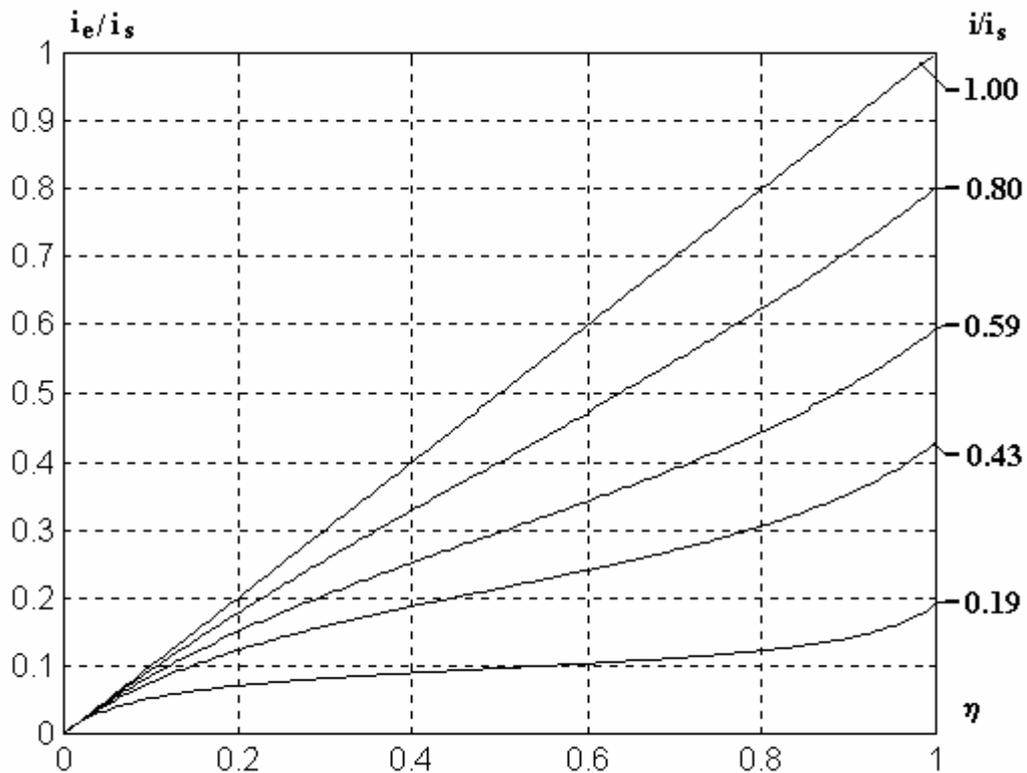

**Figure 5. Graphics of a set of solutions of (65) at different rate of recombination.**

The equation (67) is transcendent, and is solved numerically. The curve of the dependence between the current density and the electric field strength (the potential difference between both electrodes, respectively) is presented in Figure 6.

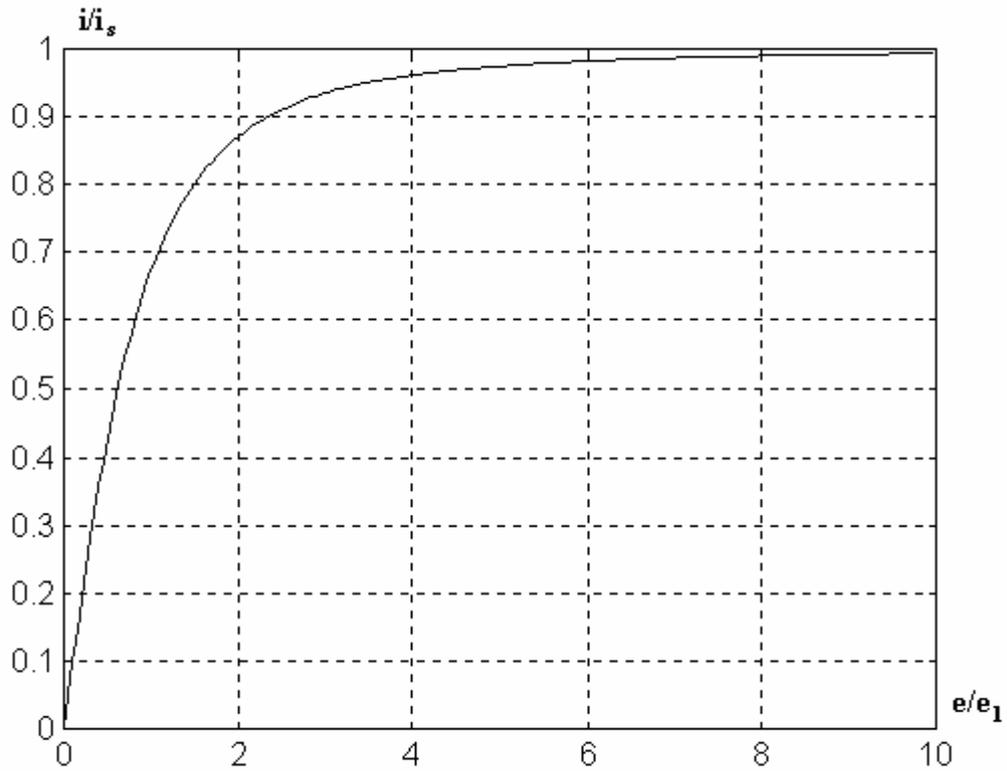

**Figure 6. A current-voltage characteristic of ionization chamber.**

As it is evident from Figure 6 with the increasing of the electric field strength the current running through the ionization chamber increases, as it goes for the current of saturation. At strong fields the current-voltage characteristic could be approximated as:

$$\frac{i}{i_s} \cong 1 - \frac{2}{3}\left(\frac{e_1}{e}\right)^2 + \frac{4}{5}\left(\frac{e_1}{e}\right)^4. \tag{68}$$

At weak fields the current-voltage characteristic could be approximately determined from (67) as:

$$\frac{i}{i_s} \cong \frac{e}{e_1}\left[1 - \frac{\pi^2}{8}\left(\frac{e}{e_1}\right)^2\right]. \tag{69}$$

## 6. Conclusion

In conclusion it could be said that a differential equation is derived which the current submits to in the volume of parallel-plane, cylindrical and spherical ionization chamber. An analytical formula of the current-voltage characteristic of an ionization chamber with homogeneous ionization is obtained in the form of a transcendent equation. For the parallel-plane case the comparison with experimental data is performed.